\documentclass[preprint,12pt]{elsarticle}

\usepackage{amsthm}

\def\|{\' \i }

\newfont{\Mb}{msbm10}

\newcommand{\dx}{{\rm d}x}
\newcommand{\dy}{{\rm d}y}

\newcommand{\Pd}[2]{{{\partial #1}\over{\partial #2}}}

\newcounter{bla}

\journal{Chaos, Solitons \& Fractals}
\usepackage[top=3.5cm, bottom=3.5cm, left=3cm, right=3cm]{geometry}

\begin{document}

\begin{frontmatter}

\title{\uppercase{The search for Invariants for 3D Systems of 1ODEs - a new Method and Integrability Analysis}}

\author[uerj]{L.G.S. Duarte}
\ead{lgsduarte@gmail.com}

\author[uerj]{J.P.C. Eiras}
\ead{joaopaulo.eiras@hotmail.com}

\author[uerj]{L.A.C.P. da Mota\corref{cor1}}
\ead{lacpdamota@uerj.br or lacpdamota@gmail.com}

\cortext[cor1]{Corresponding author  -- {\footnotesize L.G.S. Duarte and L.A.C.P. da Mota wish to thank Funda\c c\~ao de Amparo \`a Pesquisa do Estado do Rio de Janeiro (FAPERJ) for a Research Grant.}}

\address[uerj]{Universidade do Estado do Rio de Janeiro,
{\it Instituto de F\'{\i}sica, Depto. de F\'{\i}sica Te\'orica},
{\it 20559-900 Rio de Janeiro -- RJ, Brazil}} 



\begin{abstract}
In \cite{elementary3D}, we have presented the theoretical background for finding the Elementary Invariants for a 3D system of first order rational differential equations (1ODEs).  We have also provided an algorithm to find such Invariants. Here we introduce new theoretical results that will lead to a novel, more efficient, approach to determine these invariants. Furthermore, one important aspect of such dynamical systems is that the integrability can be an issue. We will show that the present theoretical development allows for an Integrability analysis in the case where the system has free parameters.
\end{abstract}


\end{frontmatter}

\section{Introduction}
\label{intro}
In \cite{elementary3D}, we have introduced a theoretical approach for finding the (possible) elementary invariants for a 3D systems of 1ODEs. The whole method and algorithms were a follow up of many results we have produced on the line of the Prelle-Singer (Darbouxian) approach \cite{PS} for solving and/or reducing ODEs \cite{Nosjpa2001,firsTHEOps1}. There has been already 6 years or so since we have completed the work presented on \cite{elementary3D} and we have produced some more work on those lines \cite{JMP,8} since then, but nothing directly related to systems. So, in a sense, the results hereby presented are long overdue. We introduce a new set of theoretical results that allows one to deal with cases where the 3D system presents a complexity that would make the ``old'' procedure unpractical. Furthermore, most importantly, we will show that we will also have a tool to analyze the Integrability of such systems and  that, obviously, constitute a powerful tool in the chaotic aspects of dynamical systems.

The paper is divided as follows: In section (\ref{elemnt}), we are going to, very briefly and brutally, summarize the results presented on \cite{elementary3D} in order to introduce some concepts and definitions that will prove essential in the explanation of the new method. Then we will introduce our new approach in section (\ref{newmethod}). The following section will present examples of the inner workings of the method and also serve as a ``poster'' for the nice features of the procedure: Such as Integrability analysis and non-elementary Invariants. In section (\ref{conclusion}), we present our conclusions and point out ways to further the ideas hereby presented.

\section{3D systems of 1ODEs with elementary invariants}
\label{elemnt}
In this section, we will introduce some basic concepts involving 3D polynomial systems of 1ODEs. These results allowed for the production of a semi-algorithm to deal with a class of 3D polynomial systems of 1ODEs presenting, at least, one elementary first integral\cite{elementary3D}. It is important that we do that summary in order to set the stage for the new theoretical results\footnote{Results to be introduced in section \ref{newmethod}} and better understand the advances we had to make to achieve it.

Consider the generic 3D system:
\begin{eqnarray}
\label{3Dsys}
\frac{dx}{dt} =  \dot{x} = f(x,y,z), \frac{dy}{dt} =  \dot{y} = g(x,y,z), \frac{dz}{dt}  = \dot{z} = h(x,y,z).
\end{eqnarray}
where $f,\,g$ and $h$ are all polynomials in $(x,y,z)$.

A function $I(x,y,z)$ is a {\it differential invariant} of the system (\ref{3Dsys})
if $I(x,y,z)$ is constant over all solution curves of (\ref{3Dsys}), i.e., $\frac{dI}{dt}=0$.

Thus, over the solutions, one has:
\begin{equation}
\label{d_inv2}
\frac{dI}{dt} = \partial_x I\, \dot{x} + \partial_y I\, \dot{y} + \partial_z I\, \dot{z} =
f\,\partial_x I + g\,\partial_y I + h\,\partial_z I = 0.
\end{equation}
\noindent
Defining the {\it Darboux operator} (see, equation (\ref{eq_def_D2})) associated with (\ref{3Dsys}),
we can write the condition for a function $I(x,y,z)$ to be a first integral
of the 3D system (\ref{3Dsys}) as $D[I]=0$, finally leading to equation (\ref{eq_def_D2}) below:
\begin{equation}
\label{eq_def_D2}
D \equiv f \,\partial_x + g \,\partial_y\, + h \,\partial_z\, \hbox{where}\,\,\, dt = \frac{dx}{f} = \frac{dy}{g} = \frac{dz}{h}
\end{equation}
\noindent
%
where $f, g$ and $h$ are the polynomials introduced in equation (\ref{3Dsys})
So, by defining the 1-forms $\alpha$ and $\beta$ as $\alpha \equiv g\,dx - f\,dy$ and $\beta \equiv h\,dx - f\,dz$, one can see that they are
null over the solutions of the system. So, one has\footnote{Please see \cite{elementary3D}, for details.}:
\begin{equation}
\label{aeb}
\alpha \equiv g\,dx - f\,dy = 0\,\,{\rm and}\,\,\beta \equiv h\,dx - f\,dz = 0. \Rightarrow \,\, dI = r\, \alpha + s\, \beta
\end{equation}
%
%
From (\ref{aeb}) we have
\begin{eqnarray}
\label{IxIyIz}
D[I] &=& dI = r \, (g\,dx - f\,dy) + s \, (h\,dx - f\,dz) \Rightarrow \nonumber \\
I_x & = & -\,r\,g - s\,h, I_y  =  r\,f, I_z  =  s\,f
\end{eqnarray}
\noindent
implying that:
\begin{eqnarray}
\label{III}
I(x,y,z) = \int (-\,r\,g - s\,h)\,\dx+\int\!\left( r\,f - \Pd{}{y}\int (-\,r\,g - s\,h) \,\dx\right)\dy + \nonumber \\
\int\!\left\{ s\,f - \Pd{}{z}\left[ \int (-\,r\,g - s\,h)\,\dx+\int\!\left( r\,f -
\Pd{}{y}\int (-\,r\,g - s\,h) \,\dx\right)\dy   \right] \right\} dz\,.
\end{eqnarray}

In order to solve (\ref{III}), one needs to determine $r$ and $s$. Very briefly, we will display the following equations that are two of the pillars where the whole method stands (both the ``old'' and the one to be presented here):
\begin{equation}
\label{PDr3}
f\,P\,\frac{D[T]}{T} = f\,P\,\sum n_i\,q_i = - f\,D[P] - P\,(f\,f_x+g\,f_y+f\,h_z) - Q\,(f\,g_z-g\,f_z) \,,
\end{equation}
\begin{equation}
\label{QDr3}
f\,Q\,\frac{D[T]}{T} = f\,Q\,\sum n_i\,q_i = - f\,D[Q] - Q\,(f\,f_x+f\,g_y+h\,f_z) - P\,(f\,h_y-h\,f_y) \,.
\end{equation}
\noindent
where $r/Q = T=\prod_i {p_i}^{n_i}$ and $s=r\frac{P}{Q}$.
Basically, these equations came from the use of the compatibility conditions $(I_{xy}=I_{yx},\,I_{xz}=I_{zx},\,I_{yz}=I_{zy} )$
and (\ref{IxIyIz}). Solving the equations above, $r$ and $s$ would have been found, we can use (\ref{III}) to obtain the first integral $I(x,y,z)$ by quadratures.



\section{New theoretical results for 3D systems of 1ODEs}
\label{newmethod}

In this section, we are going to produce a new method to find the differential invariant for the 3D system. It will prove to be more efficient for  many cases, thus broadening the scope of the method.

 In \cite{arxiv}, we have (among other things) presented a new approach to reduce 2ODEs, finding the correspondent first order differential invariant via the use of the $S$-function (defined in \cite{Nosjpa2001}). We are not going to repeat the demonstrations and theorems thereby developed. Sometimes we are going to cite these results. The results we intend to present here in more detail are the extensions of these ideas for the case of 3D rational differential systems, the new elements pertaining only systems. Consider the following differential operators:
\begin{eqnarray}
\label{dos}
D_o & \equiv & N \,\partial_x + z\, N \,\partial_y\, + M \,\partial_z \nonumber \\
D_s & \equiv & f \,\partial_x + g \,\partial_y\, + h \,\partial_z
\end{eqnarray}
\noindent
where the second equation is our previously defined operator (\ref{eq_def_D2}).
\bigskip

The first equation is the analogue of (\ref{eq_def_D2}) for the case of a rational 2ODE. What are the binding points linking these two operators? If $I_o$ and $I_s$ are the differential invariants for the 2ODE and the 3D system respectively, both operators would produce the result:
\begin{eqnarray}
\label{dos2}
D_o[I_o] & \equiv & N \,\partial_x[I_o] + z\, N \,\partial_y[I_o]\, + M \,\partial_z[I_o] = 0 \nonumber \\
D_s[I_s] & \equiv & f \,\partial_x[I_s] + g \,\partial_y[I_s]\, + h \,\partial_z[I_s] = 0
\end{eqnarray}
\noindent
So they both have in common that they ``define'' differential invariants.
\bigskip

Let us push this result a little further. Let us suppose that formally these invariants are the same. This will produce the following stream of results:
\begin{eqnarray}
\label{dos3}
I_o = I_s = I &\Rightarrow& \nonumber \\
\left( \frac{N}{f}\right) \,\,\, D_s[I]  = 0 &\Rightarrow&  N \,\partial_x[I_s] + N\frac{g}{f} \,\partial_y[I]\, + N\frac{h}{f} \,\partial_z[I] = 0 \Rightarrow \nonumber \\
\left( \frac{N}{f}\right) \,\,\, D_s[I] - D_o[I] &=&   \left( N\frac{g}{f} - z N \right)\,\partial_y[I] + \left( N\frac{h}{f} - M \right)\partial_z[I] = 0 \nonumber \\
&\Rightarrow&  N \left(\frac{g-zf}{f} \right)\,\partial_y[I] + \left( \frac{Nh - fM}{f} \right)\partial_z[I] = 0 \nonumber \\
\hbox{thus} &\Rightarrow& \left( \frac{g-zf}{f} \right) \left( N \,\partial_y[I] + \left( \frac{Nh - fM}{g-zf} \right)\partial_z[I] \right)  = 0 \nonumber \\
\hbox{and finally}& \Rightarrow & N \,\partial_y[I] + \left( \frac{Nh - fM}{g-zf} \right)\partial_z[I]  = 0
\end{eqnarray}
\noindent
Surely, the reader might be asking why all this manipulation to generate an operator that only has two derivatives, $\partial_y$ and $\partial_z$, etc. We hope to let it clear in what follows.

In \cite{arxiv}, we have developed an entire set of results and theorems that link our $S$-function (please see \cite{Nosjpa2001}) with finding the invariant for a given 2ODE.  The $S$-function we have introduced in \cite{Nosjpa2001} is now called $S_1$ and (as previously) is defined by:
\begin{eqnarray}
\label{S1}
S = S_1 &=& - \frac{\partial_y[I]}{\partial_z[I]} \Rightarrow \hbox{using this into equation (\ref{dos3})} \\
& \Rightarrow & - N \, S_1 \partial_z[I] + \left( \frac{Nh - fM}{g-zf} \right)\partial_z[I]  = 0 \nonumber \\
& \Rightarrow & \left( - N \, S_1  + \left( \frac{Nh - fM}{g-zf} \right) \right) \partial_z[I]  = 0 \nonumber \\
& \Rightarrow &  \left( - N \, S_1 \left( g-zf \right) + \left( Nh - fM \right) \right) \partial_z[I]  = 0
\end{eqnarray}
\noindent
The chain of manipulations and results so far imply that $g-zf \ne 0$ and $f \ne 0$.

In \cite{arxiv}, we have dealt with the case (that proved to be very efficient) $ S_1=\frac{P}{N}$, where $P$ is a polynomial and $N$ is the numerator of the 2ODE under study. This case proved to be of a large applicability so we are going to concentrate on that. Using $ S_1=\frac{P}{N}$ in equation (\ref{S1}), one gets:
\begin{eqnarray}
\label{S12}
& \Rightarrow &  \left( - N \, S_1 \left( g-zf \right) + \left( Nh - fM \right) \right) \partial_z[I]  = 0 \Rightarrow \nonumber \\
&\Rightarrow&   \left( - P \left( g-zf \right) + \left( Nh - fM \right) \right) \partial_z[I]  = 0 \nonumber \\
&\Rightarrow&  - P \left( g-zf \right) + \left( Nh - fM \right) =0, \hbox{\,\,\,\,\,  (if $\partial_z[I] \ne 0$)}
\end{eqnarray}
\noindent
From this, we will start to determine the $S$-function and, continuing with the steps of the method, find the desired invariant.

Consider that, in (\ref{S12}), we have 6 polynomials $M,N,P,f,g$ and $h$. Three we know ($f,g$ and $h$) and three ($M,N$ and $P$) remain to be determined. If we construct three generic polynomials and substitute them into (\ref{S12}) the resulting equation will be linear on the coefficients of those three polynomials to be determined. Therefore, much simpler to be solved than the previous method we have produced on \cite{elementary3D}.

So, that is an easier route to determining the $S$-function. Now we have ( actually) some information regarding the $S$-function. The equations above are not enough, in the case of the 3D system, to pin point completely the $S$-function. We have to run the findings of the above ``algebra'' trough the equation the $S$-function has to satisfy. Although we gave been trying to keep the paper on a more ``practical''  approach, no theorems etc. Here we are going to show in more detail where this equation come from since it only ``exists'' on a MSc Thesis so far. The result we want to introduce is:

{\bf Theorem:} {\it Consider that the 3D system defined by equation (\ref{3Dsys}) has an invariant $I$ and an $S$-function associate to that 3D system through $I$. So this $S$-function, for this 3D system  and associated $D$ operator, must satisfy the following equation:}
\begin{equation}
\label{eqStheorem}
D\left[ S \right] = \left( S \right)^2  \, \frac{(f\,g_z-g\,f_z)}{f} + \left( S \right) \, \frac{g\,f_y+f\,h_z-f\,g_y-h\,f_z}{f} - \frac{(f\,h_y-h\,f_y)}{f}
\end{equation}
\begin{itemize} \item {\bf Demonstration} \end{itemize}
Please consider equations (\ref{PDr3} and \ref{QDr3}) above. After a little manipulation, they can be put in the following format:
\begin{equation}
\label{PDr3B}
- \frac{D[T]}{T} - \frac{D[P]}{P}=  \frac{(f\,f_x+g\,f_y+f\,h_z)}{f} + \frac{Q\,(f\,g_z-g\,f_z)}{f\,P} \,,
\end{equation}
\begin{equation}
\label{QDr3B}
\frac{D[T]}{T} + \frac{D[Q]}{Q} = - \frac{(f\,f_x+f\,g_y+h\,f_z)}{f} - \frac{P\,(f\,h_y-h\,f_y)}{f\,Q} \,.
\end{equation}
If we add these two equations, we get:
\begin{equation}
\label{P+Q}
\frac{D[Q]}{Q} - \frac{D[P]}{P} = \frac{g\,f_y+f\,h_z-f\,g_y-h\,f_z}{f} - \frac{P\,(f\,h_y-h\,f_y)}{f\,Q} + \frac{Q\,(f\,g_z-g\,f_z)}{f\,P}
\end{equation}
\noindent
if we multiply equation (\ref{P+Q}) by $Q/P$, we get:
\begin{equation}
\label{P+Q2}
D\left[ \frac{Q}{P} \right] = \left( \frac{Q}{P}\right)^2  \, \frac{(f\,g_z-g\,f_z)}{f} + \left( \frac{Q}{P} \right) \, \frac{g\,f_y+f\,h_z-f\,g_y-h\,f_z}{f} - \frac{(f\,h_y-h\,f_y)}{f}
\end{equation}
\noindent
and equation (\ref{eqStheorem}) is demonstrated.

Now, let us get on with the method to find the invariant. Now, since we have $S$ determined,  the following equation is ``materialized'' from (\ref{S1}):
\begin{eqnarray}
\label{diffS}
\frac{dz}{dy} &=&  - S(x,y,z)
\end{eqnarray}
and we now know the right hand side of this equation, where $x$ is consider a parameter. This equation is called the 1ode associated with \cite{elementary3D} through $I=I(x,y,z)$. But what is the relation of that equation to the invariant of the 3D system?  Let us introduce the following result:

{\bf result 1:} $I(x,y,z)=C$ is a solution to the equations (\ref{diffS}).

\bigskip
 It is easy to see that is true if one considers that the operator $D_a=\partial_y - S\,\partial_z$ annihilates the solution to (\ref{diffS}). But $D_a[I] = (\partial_y - S\,\partial_z) [I] = I_y - S\,I_z = I_y - (\frac{I_y}{I_z})\,I_z = 0 $. Thus, $I(x,y,z)=C$ is a solution to (\ref{diffS}).

\bigskip
But it is important to stress that, if we solve (\ref{diffS}) does that mean that we would have found $I(x,y,z)$? No! In (\ref{diffS}) we have considered $x$ as a parameter. We point out to the reader that any function of $x$ (only) is an invariant for the operator $D_a$ $\Rightarrow D_a[F(x)] = (\partial_y - S\,\partial_z) [F(x)] = 0$.

So, the relation of the general solution of the associated 1ode (\ref{diffS}), $H(x,y,z)=K$ to the first integral $I(x,y,z)$ is given by $I(x,y,z)={\cal F}(x,H)$, such that ${\cal F}$ satisfies:
\begin{equation}
\label{eqF}
D_x[I] = \frac{\partial {\cal F}}{\partial x} +\left(\frac{\partial H}{\partial x} + z\,\frac{\partial H}{\partial y} + \phi\,\frac{\partial H}{\partial z}\right)  \frac{\partial {\cal F}}{\partial H} = 0.
\end{equation}
where $\phi$ is the right-hand side of the 2ode associated to the 3D system as it was determined on (\ref{S12}) and (\ref{eqStheorem}). The result above was also demonstrated on \cite{arxiv} but here we are applying, modifying it to our interests here, namely the 3D systems.

So, basically, our new method to deal with 3D systems hereby presented can be summarized as follows:

\subsection{Steps of the method}
\label{steps}
\begin{enumerate}
\item In equation (\ref{S12}), we build three generic polynomials (of a certain degree, more about that soon) to be $M,N$, and $P$. Please note that $f,g$ and $h$ we know. Equating the coefficients for the various monomials we solve (\ref{S12}) for them. Heuristically, what we do is to provide a (common) degree for $M$ and $N$ equal or greater than 2 and for $P$ we use this degree minus 1. Please remember that (\ref{S12}) would be linear on the desired coefficients, so easily solvable. By default, the degree for $M$ and $N$ is set to be that of $f$ and the degree for $P$ is either this minus 1 or 1. If we do not manage to solve the equation, we increase the degrees until we do.
    \item This is not the whole game, some coefficients might be still undetermined. We have to run these results for $M,N$ and $P$ through equation (\ref{eqStheorem}). Again, if we do not solve the equation, we return to step 1.
    \item We then turn our attention to equation (\ref{diffS}). If we solve it, we would have found $H(x,y,z)=K$.
    \item Thus we produce equation (\ref{eqF}). If we solve it, we will find the invariant $I(x,y,z)= {\cal F}(x,y,z)$
\end{enumerate}

This new method seems elaborate. But, in many cases, it solves cases where the ``old'' approach, introduced in \cite{elementary3D} fails. Mainly due to high ($\ge 3$) degree for the Darboux polynomials. Next, we are going to present an example to illustrate the use of the method and its Integrability analysis capabilities.

\section{Integrability analysis and the use of the Method}
\label{examples}

We will use the well knonw Lorenz system as our example of the usage of the method to analyze its Integrability.
\begin{eqnarray}
\label{lorenz}
\frac{dx}{dt} & = & f(x,y,z) = s\,(y.x), \frac{dy}{dt}  =   g(x,y,z) = r\,x-xz-y, \frac{dz}{dt} =  h(x,y,z) = -b\,z+ xy.\nonumber\\
\end{eqnarray}
\noindent
where $s, r$ and $b$ are parameters.

Of course, as widely known, the above system is non-integrable for the parameters ``in general'', only for specific values. Our method can deal with that. Let us try to apply the method for the Lorenz system as it stands in (\ref{lorenz}), i.e., with the parameters undetermined as yet:

The first step presented on sub-section ({\ref{steps}}) is, for that system, translated to: choosing the degrees for M, N and P, starting with degM=degN =2 and degP=1, the first successful attempt is achieved with degM=degN =4 and degP=3. This is only a partial solution  to the coefficients needed (as we have pointted out in (\ref{steps})). It is a long result and would produce pages of output. We have considered it not necessary to reproduce it here. In turn, these results are used on equation (\ref{eqStheorem}), producing their own new set of equations\footnote{Again, very large amount of data, and we have considered that uninteresting to display it here.} and finally determine the coefficients still undetermined in the first step. Let us see below.
\begin{eqnarray}
\label{final solution}
 {\bf b =1},{\it m\_16}=0,{\it m\_31}=0,{\it n\_13}=0,{\it n\_14}=0,{
\it n\_19}=0,{\it n\_2}=0,{\it n\_25}=0,\nonumber\\{\it n\_26}=0,{\it n\_28}=0,{
\it n\_31}=0,{\it n\_7}=-{\it p\_16},{\it n\_8}=0,\nonumber\\{\it p\_14}=-{\it
p\_16},{\it p\_15}=0,{\it p\_16}={\it p\_16},{\it p\_17}=0,{\it p\_18}
=0,{\it p\_2}=0,{\it p\_3}=0,\nonumber\\{\it p\_4}=0,{\it p\_5}=0,{\it p\_6}=0,{
\it p\_7}=0,{\bf r=0,s=1/2} \nonumber\\
\end{eqnarray}

Please note that we have determined (see the results emphasized in bold face above) the free parameters that appear on the Lorenz system, namely $s=\frac{1}{2},r=0$ and $b=1$. So, our method can determine the regions of integrability for the parameters on the system under scrutiny. We will not dwell here on the vast range of applications of this property of the method (apart from the obvious practical one of determining the particular set of values or range of values for the parameters for a particular case). That will be further pursued elsewhere. Here we are introducing the new ``tool''.

All these conclusions regarding the integrability for the Lorenz system were obtained with the first two steps of the procedure. Let us briefly display the next steps determining the Invariant (for the integrable scenario, of course).

In step 3, we deal with the solving of the differential equation (\ref{diffS}) that, for this case is:
\begin{equation}
\frac{dz}{dy}=\frac{ -(x^2-z) y}{(x^2 z + y^2)}
\end{equation}
\noindent
leading to:
\begin{equation}
H(x,y,z) = 1/2\,{\frac {{x}^{4}-2\,{x}^{2}z-{y}^{2}}{{x}^{4}-2\,{x}^{2}z+{z}^{2}}
} = K
\end{equation}
\noindent
which finally leads to the final step of the method and determine the invariant for the integrable case:
\begin{equation}
{\it invariant}=1/2\,{\frac {{x}^{4}-2\,{x}^{2}z-{y}^{2}}{{x}^{4}-2\,{
x}^{2}z+{z}^{2}}}
\end{equation}
\noindent

\section{Conclusions}
\label{conclusion}

In this paper, we have presented an upgrade to the results, for 3D system of 1odes, introduced in \cite{elementary3D}, where the Prelle-Singer approach was extended to this type of system. The main theoretical improvement is embodied by the new method introduced in section (\ref{newmethod}), where we use the $S$-function on the core of the new approach. The $S$-function was first introduced on \cite{Nosjpa2001} and later on used on many different ways by us and other people \cite{royal,LakshmananRajasekar,devianos citar,noscita,outroquecita}. In \cite{arxiv}, we have produced a new approach to deal with reducing 2odes also using the $S$-function as the corner stone. Here, we are presenting the counterpart method for the 3D systems. The motivation for the introduction of the new method to deal with 3D systems is that, for the cases where the necessary Darboux polynomials to use the ``old'' procedure are of too high a value ($\ge 3$), the method is unpractical for some cases. Furthermore, and perhaps more interestingly, it is possible to do an integrability analysis using our new approach, as it is shown in section (\ref{newmethod}). This may prove vital for the case of any system that presents pockets of integrability (or even single points spread over the parameter space). Our method provides that opportunity for the research community combined with the solving (determination of the Integral Invariant) of the system. Another comment we would like to make is that our present method can deal with cases where the Invariant is Liouvillian (and  ``non-elementary''). In \cite{elementary3D}, that was not the case. So, apart from the practical applicability to cases where the method introduced in \cite{elementary3D} might have run into problems, we can deal, in principle, with cases out of the scope of the previous method (only elementary Invariants). In order to further our work hereby presented, we intent to analyze the question of the ``related'' 2ode, i.e., the 2ode that presents the same Invariant as the 3D-system and the study of the pockets of Integrability.
\bibliographystyle{elsarticle-num}

\end{document}